quant-ph/0407006
\documentclass[pra,12pt,eqsecnum]{revtex4}
\usepackage{graphicx}

\begin{document}

\title{Polarization squeezing in vertical-cavity surface-emitting lasers}
\author{Yu.~M.~Golubev$^{1}$, T.~Yu.~Golubeva$^{1}$, M.~I.~Kolobov$^{2}$,
and E.~Giacobino$^{3}$}
\address{1 Physics Institute, St.~Petersburg State University,\\
198904 Petrodvorets, St.~Petersburg, Russia}
\address{2 Laboratoire PhLAM, Universit\'e de Lille 1,
F-59655 Villeneuve d'Ascq Cedex, France}
\address{3 Laboratoire Kastler Brossel, Universit\'{e} Pierre et Marie Curie,
F-75252\\ Paris Cedex 05, France}
\date{\today}

\begin{abstract}
We further elaborate the theory of quantum fluctuations in vertical-cavity
surface-emitting lasers (VCSELs), developed in Ref.~\cite{Hermier02}. In
particular, we introduce the quantum Stokes parameters to describe the
quantum self- and cross-correlations between two polarization components
of the electromagnetic field generated by this type of lasers. We
calculate analytically the fluctuation spectra of these parameters and
discuss experiments in which they can be measured. We demonstrate that in
certain situations VCSELs can exhibit polarization squeezing over some
range of spectral frequencies. This polarization squeezing has its origin
in sub-Poissonian pumping statistics of the active laser medium.
\end{abstract}

\maketitle
\newpage

\section{Introduction}

In the last years there has been an increasing interest to the
polarization properties of the vertical-cavity surface-emitting lasers
(VCSELs). This interest is motivated in the first line by the potential
applications of this type of lasers in the high-rate optical
communications~\cite{Schnitzer99}. But there is also more fundamental
reason for understanding of the polarization behavior in VCSELs, namely, a
possibility of generating the intensity-squeezed light using the
sub-Poissonian pumping of the active medium~\cite{Golubev,Yamamoto86}. To
date, squeezing in VCSELs has been demonstrated experimentally for both
single-mode operation and in a multi-transverse-mode
regime~\cite{Degen,Hermiera}. In single-mode operation with only one
linearly polarized mode above threshold the fluctuations in a
sub-threshold mode with polarization orthogonal to the lasing mode can
present large intensity noise~\cite{vanExter,Willemsen} and, moreover, be
highly correlated with the intensity fluctuations of the oscillating mode.
This phenomenon can result in deterioration of squeezing observed in
experiments with polarization sensitive optical elements. Therefore,
polarization dynamics in VCSELs plays an important role for correct
description of their quantum fluctuations.

At present, the standard theory that accounts for dynamics of two
polarization components of the electromagnetic field in VCSELs is the
so-called ``spin-flip'' model developed by San Miguel, Feng and
Moloney~\cite{SanMiguel}. On the basis of this model several authors have
formulated semiclassical theories of light fluctuations in
VCSELs~\cite{vanExter,Willemsen,vanExter98,vanExter99,Mulet01}. However,
semiclassical description is inappropriate for intensity-squeezed light
and, therefore, calls for fully quantum model of quantum fluctuations in
VCSELs. The ``quantum spin-flip'' model was developed recently in
Ref.~\cite{Hermier02}. This model takes into account, on the one side, the
dynamics of two polarization components of the electromagnetic field and,
on the other side, the pumping statistics of the active laser medium. In
particular, the quantum spin-flip theory allows for sub-Poissonian pumping
statistics in which case VCSELs can generate the intensity-squeezed light.

In this paper we further elaborate the quantum spin-flip model of VCSELs,
developed in~\cite{Hermier02}. In particular, we apply the quantum Stokes
parameters to describe the quantum self- and cross-correlations of two
polarization components of the electromagnetic field generated by VCSELs.
We analytically calculate the fluctuation spectra of the quantum Stokes
parameters and discuss experiments in which they can be measured. We
demonstrate that for the sub-Poissonian pumping statistics VCSELs can
exhibit polarization squeezing in some range of spectral frequencies.

The paper is organized as follows. In Sec.~\ref{model} we give a short
resume of the quantum spin-flip model developed in Ref.~\cite{Hermier02},
and calculate analytically the spectral densities of quantum fluctuations
of the quadrature components. In Sec.~\ref{stokes} we introduce the
quantum Stokes parameters, their fluctuation spectra, and discuss the
experiments where these fluctuations spectra can be measured. Using the
results obtained in Sec.~\ref{model} we analytically calculate the
fluctuation spectra of the quantum Stokes parameters. In
Sec.~\ref{pol_states} with help of the analytical results obtained in
Sec.~\ref{stokes} we illustrate graphically the possibilities of
observation of polarization squeezing in VCSELs. We also provide the
figures of typical cross-correlations spectra of photocurrents and
cross-correlation spectra of the Stokes parameters $S_2$ and $S_3$ that
can be measured experimentally. In Sec.~\ref{concl} we summarize the
results.

\section{Quantum spin-flip theory of VCSELs}
\label{model}
\subsection{Resume of the model}

In this section we shall give a brief resume of the quantum spin-flip
model of VCSELs developed in Ref.~\cite{Hermier02}. We shall define the
physical parameters of this model and provide the equations which will be
used in the following sections. For more details we refer the reader to
Ref.~\cite{Hermier02}.

The semiclassical four-level spin-flip model of VCSELs was developed by
San Miguel, Feng and Moloney \cite{SanMiguel}. This model describes very
well the dynamics of these semiconductor lasers and is widely used for
understanding of such phenomena, for example, as polarization switching.
The spin-flip model takes into account the spin sublevels of the total
angular momentum of the heavy holes in the valence band and of the
electrons in the conduction band. These four sublevels interact with two
circularly polarized electromagnetic waves in the laser resonator and it
is this interaction that is responsible for the complicated polarization
dynamics manifested by this type of lasers.

The four-level scheme of the semiconductor medium is shown in Fig.~1. Two
lower levels $|b\pm\rangle$ correspond to the unexcited state of the
semiconductor medium with zero electron-hole pairs while the upper levels
$|a\pm\rangle$ to the excited states with an electron-hole pair
created~\cite{Koch}. Two pairs of levels $|a+\rangle, |b+\rangle$ and
$|a-\rangle, |b-\rangle$ are coupled via interaction with the left and
right circularly polarized electromagnetic waves in the laser cavity
described by the field operators $\hat{a}_+(t)$ and $\hat{a}_-(t)$. As
explained in Ref.~\cite{SanMiguel}, physically these two pairs of
transitions are associated with two $z$-components $J_z=\pm 1/2$ of the
total angular momentum $J=1/2$ of the electrons in the conduction band and
corresponding $z$-components $J_z=\pm 3/2$ for $J=3/2$ of the heavy holes
in the valence band. The constants $\gamma_a$ and $\gamma_b$ are the decay
rates of the populations of the upper and lower levels, $\gamma_{\perp}$
(not shown in Fig.~1) is the decay rate of the polarization, and
$\gamma_c$ is the spin-flip rate that accounts for mixing of populations
with opposite values of $J_z$. The last parameter was introduced in
Ref.~\cite{SanMiguel} to describe the spin-flip relaxation process. This
parameter is responsible for coupling of two transitions with different
circular polarizations and, as a result, for various polarization dynamics
of VCSELs.

It should be noted that the authors of Ref.~\cite{SanMiguel} have
considered the situations of equal relaxation constants of the upper and
the lower levels, $\gamma_a=\gamma_b$. However, it is known from the
literature~\cite{Golubev,Yamamoto86} that this is not the most favorable
condition for generation of the sub-Poissonian light. Therefore, the
quantum spin-flip theory in Ref.~\cite{Hermier02} was developed for
arbitrary values of $\gamma_a$ and $\gamma_b$. In this paper we shall also
consider this general situation.

Moreover, it has been mentioned in the literature (see, for example,
Ref.~\cite{Hermier02}) that this model describes correctly a semiconductor
laser if we assume the decay rate $\gamma_b$ of the lower levels to be
very large compared to the other decay constants, namely, $\gamma_a,
\gamma_c$ and $\kappa$. From the classical point of view both situations
$\gamma_b=\gamma_a$ and $\gamma_b\gg\gamma_a$ result in the same dynamical
behavior of VCSELs. However, it turns out that the statistical properties
of two models with $\gamma_b=\gamma_a$ and $\gamma_b\gg\gamma_a$ are very
different. The detailed discussion of this difference is out of the scope
of this paper and we shall address this point elsewhere.

We have indicated in Fig.~1 the pump process with mean total pumping rate
$2R$ which is then separated with equal probabilities between two
sublevels $|a_+\rangle$ and $|a_-\rangle$. Quantum spin-flip model of
Ref.~\cite{Hermier02} takes into account a possibility of sub-Poissonian
pumping of the laser medium using the technique of the pump-noise
suppression~\cite{Golubev,Yamamoto86}. For stationary in time average
pumping rate, the influence of the pump statistics can be characterized by
a single parameter $p\leq1$~\cite{Benkert,Kolobov}. For $p=1$ the pump is
perfectly regular while for $p=0$ the pump has Poissonian statistics.
Intermediate values of $0\leq p\leq 1$ correspond to sub-Poissonian
pumping while for $p\leq 0$ the pump process possess the excess classical
fluctuations and corresponds to super-Poissonian statistics.

This pump statistics was introduced into the quantum spin-flip model using
the Heisenberg-Langevin equations for the operator-valued collective
populations ${\hat N}_{a\pm}(t)$, ${\hat N}_{b\pm}(t)$ of the upper and
lower levels in Fig.~1, and for the collective polarization ${\hat
P}_{\pm}(t)$. On the basis of the Heisenberg-Langevin equations, the
equivalent $c$-number Langevin equations were derived for the collective
atomic and field variables, corresponding to the normal ordering of the
atomic and field operators~\cite{Benkert,Kolobov}. Next, using the fact
that the relaxation rates $\gamma_b$ of the lower levels and
$\gamma_{\perp}$ of the polarization in VCSELs are much bigger than the
relaxation rate $\gamma_a$ of the upper levels, the macroscopic $c$-number
populations $N_{b\pm}(t)$ and the macroscopic $c$-number polarization
$P_{\pm}(t)$ were adiabatically eliminated. The resulting equations can be
written in terms of the total population of two upper levels $|a+\rangle$
and $|a-\rangle$, and of the total inversion between them. The
corresponding variables are defined as $D(t)=(N_{a+}(t)+N_{a-}(t))/2$, and
$d(t)=(N_{a+}(t)-N_{a-}(t))/2$. The equations for these variables and the
two $c$-number field components $a_{\pm}(t)$ are
\begin{equation}
     \dot{a}_{\pm}(t)=-\kappa a_{\pm}(t)-(\kappa_a+i\omega_{p})a_{\mp}(t)
     +c(1-i\alpha )\Bigl[D(t)\pm d(t)\Bigr]a_{\pm}(t)
     +F_{\pm}(t),
               \label{SFMes1}
\end{equation}
\begin{equation}
      \dot{D}(t)= R-\gamma D(t)-c(|a_{+}(t)|^2+|a_{-}
      (t)|^2)D(t)-c(|a_{+}(t)|^2-|a_{-}(t)|^2)d(t)
      +F_{D}(t),
                \label{SFMes2}
\end{equation}
\begin{equation}
      \dot{d}(t)=-\gamma _{s}d(t)-c(|a_{+}(t)|^2-|a_{-}
      (t)|^2)D(t)-c(|a_{+}(t)|^2+|a_{-}(t)|^2)d(t)
      +F_{d}(t).
                \label{SFMes3}
\end{equation}
Here $\kappa$ is the cavity damping constant, $\omega_p$ and $\kappa_a$
describe the linear birefringence and the linear dichroism of the
semiconductor medium. The last parameter was not included into the model
in Ref.~\cite{Hermier02} and is introduced here as a generalization. Next,
$\alpha$ is the linewidth enhancement in semiconductor lasers,
\begin{equation}
     \alpha=\frac{\nu-\omega}{\gamma_{\perp}},
\end{equation}
where $\nu$ is the frequency of the semiconductor energy gap, and $\omega$
is the resonator frequency. We have also defined the relaxation rate
$\gamma_s$ as $\gamma_s=\gamma_a+2\gamma_c$, and have introduced the
following shorthands,
\begin{equation}
      c=\frac{g^2}{\gamma_{\perp}(1+\alpha^2)}, \hspace{5mm}
      \gamma=\gamma_a,
                 \label{shorthand}
\end{equation}
where $g$ is the coupling constant of interaction of the electromagnetic field
with the polarization.

The functions $F_{\pm}(t)$, $F_D(t)$ and $F_d(t)$ are the $c$-number Langevin
forces. Their nonzero correlation functions were calculated in
Ref.~\cite{Hermier02}. In general the results are rather cumbersome but they are
simplified in the vicinity of the stationary solutions. For completeness we
shall give these correlation functions for the stationary solutions at the end
of this section.

\subsection{Stationary semiclassical solutions}

Semiclassical equations of VCSELs are obtained from
Eqs.~(\ref{SFMes1})-(\ref{SFMes3}) by dropping the $c$-number Langevin
forces. In this subsection we shall give the stationary solutions of these
equations which characterize the stationary generation of VCSELs. For
investigation of quantum fluctuations in VCSELs we shall use standard
assumption that these fluctuations are small compared to the corresponding
stationary values. This will allow for linearization of
Eqs.~(\ref{SFMes1})-(\ref{SFMes3}) around stationary solutions with
respect to the quantum fluctuations.

Stationary solutions of Eqs.~(\ref{SFMes1})-(\ref{SFMes3}) have been
investigated in detail in~\cite{Martin-Regalado97,SanMiguel}. When
$\omega_p\neq 0$ and $\kappa_a\neq 0$ there are in general four types of
stationary solutions: two of them have linear polarization along the $x$
and $y$ axes, and two other elliptical polarization. We shall consider
only linearly polarized solutions because this type of solutions is
usually realized in experiments. In this case the circularly polarized
field components have equal amplitudes and can be written in the form
\begin{equation}
     a_{\pm}(t)=Qe^{i(\Delta t\pm\psi)},
              \label{stat}
\end{equation}
where the real amplitude $Q$ is normalized so that $Q^2=|a_+|^2=|a_-|^2$
gives the mean number of photons in the corresponding circularly polarized
field mode. Two other parameters $\Delta$ and $\psi$ determine the type of
polarization of the stationary solution (\ref{stat}).

We remind that the linearly polarized field components $a_x(t)$ and
$a_y(t)$ are related to the circularly polarized ones as
\begin{equation}
     a_{x}(t)=\frac{a_{+}(t)+a_{-}(t)}{\sqrt{2}}, \quad
     a_{y}(t)=\frac{a_{+}(t)-a_{-}(t)}{\sqrt{2}i}.
               \label{linearpolariz}
\end{equation}
For the $x$-polarized solution $\psi=0$ and for the $y$-polarized solution
$\psi=\pi/2$. The frequency detunings $\Delta$ in Eq.~(\ref{stat}) are
different for these solutions and are equal to
\begin{equation}
      \Delta_{x,y}=-[\kappa_{x,y}\alpha\pm\omega_{p}],
\end{equation}
where the upper sign corresponds to the $x$-polarized solution and the
lower sign to the $y$-polarized one. Here we have introduced the
shorthands $\kappa_x=\kappa+\kappa_a$ and $\kappa_y=\kappa-\kappa_a$. The
$x$-polarized stationary solution reads
\begin{equation}
     a_x=\sqrt{2}Qe^{i\Delta_x t}, \quad a_y=0,
            \label{statx}
\end{equation}
while the $y$-polarized stationary solution is given by
\begin{equation}
     a_x=0, \quad a_y=\sqrt{2}Qe^{i\Delta_y t}.
\end{equation}
For both solutions we have
\begin{equation}
     Q=\sqrt{I_s(r-1)},
           \label{stationary}
\end{equation}
where $r=R/R_{\rm th}$ is the dimensionless pumping rate, $R_{\rm th}$ is the
threshold pumping rate, and $I_s$ is the saturation intensity; the two latter
are given by
\begin{equation}
     R_{\rm th}=\frac{\gamma\kappa_{x,y}}{c}, \quad
     I_s=\frac{\gamma}{2c}.
\end{equation}
Note that for $\kappa_a > 0$ the threshold pumping rate for the
$y$-polarized solution is lower that for the $x$-polarized one.

The stationary values of the atomic variables $d_0$ and $D_0$ for these
linearly polarized solutions are equal to
\begin{equation}
     d_0=0,\quad D_0=\frac{R}{\gamma+2cQ^2}=\frac{\kappa_{x,y}}{c}.
            \label{stat_atomic}
\end{equation}
In the case of VCSELs as in general for solid-state and semiconductor
lasers the question of stability of stationary solutions is very
important. The stability analysis of these stationary solutions was
performed in a number of publications, as for example,
Refs.~\cite{Martin-Regalado97, Golubev03}, and we refer the reader to this
papers for details. In our analysis of quantum fluctuations we shall
assume that the corresponding stationary operation regime of VCSEL is
stable. Since for low pumping rate only $x$-polarized solution is stable,
we shall restrict our analysis of quantum fluctuations only for this type
of stationary solutions.

\subsection{Linearization around stationary solutions}

To calculate the quantum fluctuations around the stationary solution we
shall linearize Eqs.~(\ref{SFMes1})-(\ref{SFMes3}) around the steady state
given by Eq.~(\ref{stat}). As mentioned above we shall consider here only
$x$-polarized stationary solution. Adding small fluctuations to the
stationary solutions we can write the field and the atomic variables as
\begin{equation}
     a_{\pm}(t)=(Q+\delta a_{\pm}(t))e^{i\Delta t}, \quad
     D(t)=D_0+\delta D(t), \quad
     d(t)=\delta d(t).
\end{equation}
In this equation and in what follows we have dropped the index $x$ in $\Delta_x$
since we shall be concerned only with $x$-polarized solution.
Substituting these expressions into Eqs.~(\ref{SFMes1})-(\ref{SFMes3}) and
linearizing, we arrive at the following equations for small fluctuations,
\begin{eqnarray}
     \frac{d}{dt}\delta a_{\pm}(t)&=&(\kappa_a+i\omega_p)\Bigl(\delta a_{\pm}(t)-
     \delta a_{\mp}(t)\Bigr)+c(1-i\alpha)Q(\delta D(t)\pm
     \delta d(t))+F_{\pm}(t)e^{-i\Delta t},
              \nonumber\\
     \frac{d}{dt}\delta D(t)&=&-\left(\gamma+2cQ^2\right)
     \delta D(t)-\kappa_x Q\left(\delta a_{+}(t)+\delta a_{-}(t)+c.c.\right)
     +F_{D}(t),
              \nonumber\\
     \frac{d}{dt}\delta d(t)&=&-\left(\gamma_{s}+2cQ^2\right)
     \delta d(t)-\kappa_x Q\left(\delta a_{+}(t)-\delta a_{-}(t)+c.c.\right)
     +F_{d}(t).
               \label{fluctsys}
\end{eqnarray}
It is convenient to introduce the fluctuations of the linearly polarized
components of the field $\delta a_x(t)$ and $\delta a_y(t)$, defined
according to Eq.~(\ref{linearpolariz}), for which the set of coupled
equations (\ref{fluctsys}) decouples in two sets of independent equations
for $\delta a_x(t)$ and $\delta a_y(t)$ with Langevin forces $F_x(t)$ and
$F_y(t)$ defined similar to Eq.~(\ref{linearpolariz}). Moreover, we shall
define the fluctuations of the amplitude and the phase quadrature
components, $\delta X_x(t)$ and $\delta Y_x(t)$ of the $x$-polarized field
component,
\begin{equation}
      \delta X_x(t)=\frac{1}{2}\Bigl(\delta a_{x}(t)+\delta a_{x}
      ^{\ast}(t)\Bigr),\quad
      \delta Y_x(t)=\frac{1}{2i}\Bigl(\delta a_{x}(t)-\delta a_{x}
      ^{\ast}(t)\Bigr),
           \label{quadratures}
\end{equation}
and similar for the $y$-polarized component. For these fluctuations we
obtain the following equations,
\begin{eqnarray}
     \frac{d}{dt}\delta X_{x}(t)&=&\sqrt{2}cQ\delta D(t)+R_{x}(t),
              \nonumber\\
     \frac{d}{dt}\delta Y_{x}(t)&=&-\sqrt{2}\alpha cQ\delta D(t)+T_{x}(t),
              \nonumber\\
     \frac{d}{dt}\delta D(t)&=&-\Gamma\delta D(t)-
     2\sqrt{2}\kappa_x Q\delta X_{x}(t)+F_{D}(t),
              \label{polxtime}
\end{eqnarray}
and
\begin{eqnarray}
     \frac{d}{dt}\delta X_{y}(t)&=&2\kappa_a\delta X_{y}(t)
     -2\omega_p \delta Y_{y}(t)-
     \sqrt{2}\alpha cQ\delta d(t)+R_{y}(t),
              \nonumber\\
     \frac{d}{dt}\delta Y_{y}(t)&=&2\kappa_a\delta Y_{y}(t)
     +2\omega_p \delta X_{y}(t)-
     \sqrt{2} cQ\delta d(t)+T_{y}(t),
              \nonumber\\
     \frac{d}{dt}\delta d(t)&=&-\Gamma_s\delta d(t)+
     2\sqrt{2}\kappa_x Q\delta Y_{y}(t)+F_{d}(t),
              \label{polytime}
\end{eqnarray}
where the new Langevin forces $R_x(t)$ and $S_x(t)$ are defined as
\begin{eqnarray}
      R_x(t)&=&\frac{1}{2}\Bigl(F_{x}(t)e^{-i\Delta t}+F_{x}
      ^{\ast}(t)e^{i\Delta t}\Bigr),\quad
      T_x(t)=\frac{1}{2i}\Bigl(F_{x}(t)e^{-i\Delta t}-F_{x}
      ^{\ast}(t)e^{i\Delta t}\Bigr),
              \nonumber \\
      R_y(t)&=&\frac{1}{2}\Bigl(F_{y}(t)e^{-i\Delta t}+F_{y}
      ^{\ast}(t)e^{i\Delta t}\Bigr),\quad
      T_y(t)=\frac{1}{2i}\Bigl(F_{y}(t)e^{-i\Delta t}-F_{y}
      ^{\ast}(t)e^{i\Delta t}\Bigr).
\end{eqnarray}
In Eqs.~(\ref{polxtime}) and (\ref{polytime}) we have introduced
\begin{equation}
      \Gamma\equiv\gamma+2cQ^2=\gamma r,\quad
      \Gamma_s\equiv\gamma_s+2cQ^2=\gamma_s+\gamma(r-1),
\end{equation}
as convenient shorthands.

\subsection{Spectral densities of quantum fluctuations}

To solve Eqs.~(\ref{polxtime}) and (\ref{polytime}) we take the Fourier
transform of the field and atomic fluctuations,
\begin{equation}
     \delta X_{x}(\Omega)=\frac{1}{\sqrt{2\pi}}\int\limits_{-\infty }
     ^{+\infty }\delta X_{x}(t)e^{i\Omega t}dt,
             \label{fourier}
\end{equation}
and similar for the other variables, that converts these differential
equations into algebraic ones. The spectral correlation functions of these
quadratures are $\delta$-correlated,
\begin{eqnarray}
      \langle \delta X_{i}(\Omega)\delta X_{i}(\Omega')\rangle &=&
      (\delta X^2_i)_{\Omega}\delta(\Omega+\Omega'),
               \nonumber\\
      \langle \delta Y_{i}(\Omega)\delta Y_{i}(\Omega')\rangle &=&
      (\delta Y^2_i)_{\Omega}\delta(\Omega+\Omega'),
               \nonumber\\
      \langle \delta X_{i}(\Omega)\delta Y_{i}(\Omega')\rangle &=&
      (\delta X_i\delta Y_i)_{\Omega}\delta(\Omega+\Omega'),
               \label{spectrquadii}
\end{eqnarray}
with $(\delta X^2_i)_{\Omega}$, $i=x,y$ and $(\delta Y^2_i)_{\Omega}$ being
the spectral densities of the corresponding quadratures, and
$(\delta X_i\delta Y_i)_{\Omega}$ their cross-spectral density.

After a simple algebra we obtain the following expressions for the
fluctuations of the amplitude quadratures $\delta X_{x}(\Omega)$ and
$\delta X_{y}(\Omega)$, and the phase quadrature $\delta Y_{y}(\Omega)$:
\begin{equation}
      \delta X_{x}(\Omega)
      =\frac{1}{D_x(\Omega)}
      \Bigl\{(\Gamma-i\Omega)R_x(\Omega)+\sqrt{2}cQF_D(\Omega)\Bigr\},
                 \label{solx}
\end{equation}
\begin{eqnarray}
      & &\delta X_{y}(\Omega)=\frac{1}{D_y(\Omega)}
      \Bigl\{[2\kappa_x\gamma(r-1)-(2\kappa_a+i\Omega)(\Gamma_s-i\Omega)]R_y(\Omega)
      \Bigr.
              \nonumber \\
      \Bigl.
      &-&[2\alpha\kappa_x\gamma(r-1)+2\omega_p(\Gamma_s-i\Omega)]
      T_y(\Omega)+\sqrt{2}cQ(2\omega_p+2\alpha\kappa_a+i\alpha\Omega)F_d(\Omega)\Bigr\},
                 \label{solxy}
\end{eqnarray}
\begin{eqnarray}
      \delta Y_{y}(\Omega)&=&\frac{1}{D_y(\Omega)}
      \Bigl\{2\omega_p(\Gamma_s-i\Omega)R_y(\Omega)
      \Bigr.
              \nonumber \\
      \Bigl.
      &-&(2\kappa_a+i\Omega)(\Gamma_s-i\Omega)
      T_y(\Omega)+\sqrt{2}cQ(-2\alpha\omega_p+2\kappa_a+i\Omega)F_d(\Omega)\Bigr\},
                 \label{solyy}
\end{eqnarray}

with
\begin{eqnarray}
     D_x(\Omega)&=&-i\Omega(\Gamma-i\Omega)+2\kappa_x\gamma(r-1),
                    \nonumber \\
     D_y(\Omega)&=&(\Gamma_s-i\Omega)[(2\omega_p)^2+(2\kappa_a+i\Omega)^2]
     +2\kappa_x\gamma(r-1)(2\alpha\omega_p-2\kappa_a-i\Omega).
                    \label{denominatory}
\end{eqnarray}
The other phase quadrature $\delta Y_{x}(\Omega)$ will not appear in the
observables that we shall discuss below. Using the results obtained in
Ref.~\cite{Hermier02}and taking into account the stationary solutions
(\ref{stat}) and (\ref{stat_atomic}) we obtain the following nonzero
correlation functions of the Langevin forces $R_{i}(t), T_{i}(t)$ with
$i=x,y$, and $F_D(t),F_d(t)$ for the stationary regime of VCSEL in
approximation of the small fluctuations,
\begin{eqnarray}
      \langle R_x(t)R_x(t')\rangle&=&
      \langle R_y(t)R_y(t')\rangle=\langle T_x(t)T_x(t')\rangle=
      \langle T_y(t)T_y(t')\rangle=
      \kappa_x\delta(t-t'),
                \nonumber \\
      \langle F_D(t)F_D(t')\rangle&=&
      \frac{\kappa_x}{c}\Gamma\Bigl(1-\frac{1}{2}p\Bigr)\delta(t-t'),
                \nonumber \\
      \langle F_d(t)F_d(t')\rangle&=&
      \frac{\kappa_x}{c}\Gamma_s\delta(t-t'),
                \nonumber \\
      \langle F_D(t)R_x(t')\rangle&=&
      \langle F_d(t)T_y(t')\rangle=
      -\sqrt{2}\kappa_x Q\delta(t-t').
                \label{diff4}
\end{eqnarray}
Equations (\ref{solx})-(\ref{denominatory}) together with correlation
functions (\ref{diff4}) allow us to evaluate an arbitrary correlation
function of the laser light emitted by the VCSEL. The spectral densities
of the amplitude quadratures $(\delta X_x^2)_{\Omega}$, $(\delta
X_y^2)_{\Omega}$ are given by,
\begin{equation}
      (\delta X_x^2)_{\Omega}
      =\frac{\kappa_x}{|D_x(\Omega)|^2}
      \Bigl\{\Omega^2+\gamma^2r\Bigl[1-(r-1)p/2\Bigr]\Bigr\},
                 \label{X_x}
\end{equation}
\begin{equation}
      (\delta X_y^2)_{\Omega}
      =\frac{\kappa_x}{2|D_y(\Omega)|^2}
      \Bigl\{\Omega^4+A_X\Omega^2+4B_X\Bigr\},
                 \label{X_y}
\end{equation}
with $A_X$ and $B_X$ determined as,
\begin{eqnarray}
      A_X &=&
      \Bigl[2\kappa_a-\gamma(r-1)\Bigr]^2+\Bigl[2\omega_p+\alpha\gamma(r-1)\Bigr]^2-
      4\kappa\gamma(r-1)
              \nonumber \\
      &+&\gamma_s\Bigl[\gamma_s+\gamma(r-1)(\alpha^2+2)\Bigr],
              \nonumber \\
      B_X &=&
      \Bigl[\kappa_a\gamma_s-\kappa\gamma(r-1)\Bigr]^2+\Bigl[\omega_p\gamma_s+
      \alpha\kappa\gamma(r-1)\Bigr]^2+\gamma_s\gamma(r-1)(\alpha\kappa_a+\omega_p)^2,
                 \label{B_X}
\end{eqnarray}
The spectral density of the phase quadrature component $(\delta
Y_y^2)_{\Omega}$ is equal to,
\begin{equation}
      (\delta Y_y^2)_{\Omega}
      =\frac{\kappa_x}{2|D_y(\Omega)|^2}
      \Bigl\{\Omega^4+A_Y\Omega^2+4B_Y\Bigr\},
                 \label{Y_y}
\end{equation}
with $A_Y$ and $B_Y$ given by,
\begin{eqnarray}
      A_Y &=&
      4(\kappa_a^2+\omega_p^2)+\gamma_s^2+\gamma(r-1)(4\alpha\omega_p+\gamma_s),
              \nonumber \\
      B_Y &=&
      \gamma_s^2(\kappa_a^2+\omega_p^2)+\gamma_s\gamma(r-1)
      \Bigl[\omega_p^2(\alpha^2+2)+\kappa_a^2\Bigr]^2+
      \omega_p^2\gamma^2(r-1)^2(\alpha^2+1),
                 \label{B_Y}
\end{eqnarray}
Finally the cross-spectral density $(\delta X_y\delta Y_y)_{\Omega}$
reads,
\begin{equation}
      (\delta X_y\delta Y_y)_{\Omega}
      =\frac{-\kappa_x\gamma(r-1)}{2|D_y(\Omega)|^2}
      \Bigl\{\alpha\kappa_x\Omega^2+2\kappa\omega_p\gamma(r-1)(\alpha^2+1)
      +2\gamma_s\Bigl[\kappa(\alpha\kappa_a+\omega_p)+\alpha\kappa_a(\kappa_a-
      \alpha\omega_p)\Bigr]\Bigr\},
                 \label{XY_y}
\end{equation}
These analytical results will be used below for evaluation of the spectral
densities of the quantum Stokes parameters, their cross-spectral densities
and for the cross-correlation spectra of the photocurrents.

\section{Quantum polarization states of light: general discussion}
\label{stokes}
\subsection{Quantum Stokes parameters}

There are two equivalent descriptions of the polarization properties of
light in classical optics either by the polarization matrix or in terms of
the classical Stokes parameters~\cite{Born99}. During the last decade the
quantum-mechanical version of the classical Stokes parameters was
introduced in the literature and very actively used in quantum optics to
describe the quantum fluctuations of polarization of the electromagnetic
field~\cite{Jauch76,Robson74,Chirkin93,Klyshko97}. There have been several
theoretical proposals for generation of polarization-squeezed
light~\cite{Chirkin93,Korolkova94,Chirkin95,Alodjants95,Korolkova96,Korolkova02}
and a few experiments in which such kind of light was observed
~\cite{Grangier87,Karasev93,Buchev01,Bowen02}.

We shall use the language of the quantum Stokes parameters for
characterization of the quantum fluctuations of polarized light in VCSELs.
In this section we shall express the fluctuation spectra of the quantum
Stokes parameters through the spectral densities of the quadrature
components evaluated above. In the next section we shall apply these
results for the particular case of VCSELs.

Let us write the operator $\vec{\hat{E}}(t)$ of the electromagnetic field at the
output of the VCSEL in terms of the $x$- and $y$-polarized components,
\begin{equation}
     \vec{\hat{E}}(t)=\hat{a}_{x}(t)\vec{e}_{x}+\hat{a}_{y}(t)\vec{e}_{y},
           \label{vector_field}
\end{equation}
where $\hat{a}_{x}(t)$ and $\hat{a}_{y}(t)$ are the photon annihilation
operators in the Heisenberg representation. In what follows we shall omit
the time argument when this does not create ambiguities. The quantum
Stokes operators $\hat{S}_{\mu}, \mu=0,1,2,3$ are introduced similarly to
their classical counterparts (see, for example~\cite{Korolkova02}),
\begin{eqnarray}
     \hat{S}_0&=&\hat{a}^{\dag}_x\hat{a}_x+\hat{a}^{\dag}_y\hat{a}_y,
               \nonumber \\
     \hat{S}_1&=&\hat{a}^{\dag}_x\hat{a}_x-\hat{a}^{\dag}_y\hat{a}_y,
               \nonumber \\
     \hat{S}_2&=&\hat{a}^{\dag}_x\hat{a}_y+\hat{a}^{\dag}_y\hat{a}_x,
               \nonumber \\
     \hat{S}_3&=&i(\hat{a}^{\dag}_y\hat{a}_x-\hat{a}^{\dag}_x\hat{a}_y).
               \label{StokesS}
\end{eqnarray}
Using the commutation relations for the photon annihilation and creation
operators,
\begin{equation}
     [\hat{a}_i,\hat{a}^{\dag}_j]=\delta_{ij},\quad(i,j=x,y),
             \label{comm}
\end{equation}
it is easy to verify that the operator $\hat{S}_0$ commutes with all the others,
\begin{equation}
     [\hat{S}_0,\hat{S}_{\mu}]=0,\quad(\mu=1,2,3),
             \label{commS0}
\end{equation}
and that the operators $\hat{S}_1$, $\hat{S}_2$ and $\hat{S}_3$ satisfy the
commutation relations similar to the components of the angular-momentum
operator,
\begin{equation}
     [\hat{S}_1,\hat{S}_2]=2i\hat{S_3},\quad
     [\hat{S}_2,\hat{S}_3]=2i\hat{S_1},\quad
     [\hat{S}_3,\hat{S}_1]=2i\hat{S_2}.
             \label{commS123}
\end{equation}
The noncommutativity of these three Stokes operators does not allow their
simultaneous measurement in any real physical experiment. The mean values
$\langle\hat{S}_{\mu}\rangle, \mu=1,2,3$ and the variances $\Delta
S_{\mu}=\sqrt{\langle(\hat{S}_{\mu}-\langle\hat{S}_{\mu}\rangle)^2\rangle}$
are given by the uncertainty relations~\cite{Jauch76},
\begin{equation}
     \Delta S_1 \Delta S_2\geq|\langle\hat{S_3}\rangle|,\quad
     \Delta S_2 \Delta S_3\geq|\langle\hat{S_1}\rangle|,\quad
     \delta S_3 \delta S_1\geq|\langle\hat{S_2}\rangle|.
             \label{uncertainty}
\end{equation}
When the $x$- and $y$-polarized components of the electromagnetic field
are in coherent states $|\alpha_x\rangle$ and $|\alpha_y\rangle$ i.~e.,~
\begin{equation}
     \hat a_x|\alpha_x\rangle=\alpha_x|\alpha_x\rangle,\qquad\hat
     a_y|\alpha_y\rangle=\alpha_y|\alpha_y\rangle,
\end{equation}
one can speak about the {\it coherent polarization state} of the
electromagnetic field. The mean values of the quantum Stokes parameters in
this state are obtained by replacing $\hat a_x\to\alpha_x$ and $\hat
a_y\to\alpha_y$ in Eq.~(\ref{StokesS}). For example, for the first two
parameters one obtains,
\begin{eqnarray}
      &&\langle\hat S_0\rangle =|\alpha_x|^2+|\alpha_y|^2=\langle\hat n_x
      \rangle+\langle\hat n_y\rangle=\langle\hat n\rangle,
         \nonumber\\
      &&\langle\hat S_1\rangle =|\alpha_x|^2-|\alpha_y|^2=\langle
      \hat n_x\rangle-\langle\hat n_y\rangle,
\end{eqnarray}
where $\langle\hat n\rangle$ is the mean total number of photons in the
electromagnetic wave. The variances of all four quantum Stokes parameters
in this case are equal and given by~\cite{Korolkova02},
\begin{equation}
     \Delta S_\mu^2=\langle\hat n_x\rangle+\langle\hat
     n_y\rangle=\langle\hat n\rangle,\qquad\mu=0,1,2,3.
\end{equation}
This property of the coherent polarization state allows one to define a
{\it polarization squeezed state} similar to the definition of a
single-mode squeezed state. According to \cite{Chirkin93} one can speak
about polarization squeezing if one of the four variances $\Delta S_\mu$
of the Stokes parameters becomes smaller than that in the coherent state,
i.~e.~ $\Delta S_\mu^2<\langle\hat n\rangle$ for at least one $\mu$.

Classical Stokes parameters $S_{\mu}, \mu=0,1,2,3$ (without hats) are
obtained as the mean values of their quantum versions defined in
Eq.~(\ref{StokesS}), $S_{\mu}=\langle\hat{S}_{\mu}\rangle$. From the
classical point of view, all polarization properties of light are
completely described by these four parameters: $S_0$ determines the total
beam intensity, while three other parameters characterize the polarization
state of the light beam. This polarization state in classical optics is
often represented in a Poincar\'e sphere with $S_1$, $S_2$ and $S_3$
forming its three orthogonal axes.

In quantum optics to completely characterize polarization properties of
light in addition to the mean values $S_\mu$ of the quantum Stokes
parameters one has to determine their variances $\Delta S_\mu$. In general
all these variances can be different and one can speak of an uncertainty
ellipsoid in the Stokes-Poincar\'e space~\cite{Klyshko97}. In general
case, when different Stokes components are correlated, there are three
additional parameters which determine the orientation axes of this
uncertainty ellipsoid.

While the general description is outside of the scope of our paper, we
shall illustrate below graphically that in the case of VCSELs different
quantum Stokes components $\hat{S}_{\mu}$ can have different variances
$\Delta S_{\mu}$. The quantum fluctuations of polarization in VCSELs are
therefore characterized by an uncertainty ellipsoid in the
Stokes-Poincar\'e space.

\subsection{Measurement of the classical Stokes parameters}

Four classical Stokes parameters $S_{\mu}$ can be measured  in an
experimental setup shown in Fig.~2. This measurement scheme consists of a
compensator, a polarizing beam splitter (PBS), and two photodetectors. Let
$\delta_x$ and $\delta_y$ denote the phase changes produced by the
compensator in the $x$- and $y$-components of the electromagnetic field
given by Eq.~(\ref{vector_field}). Next, let $\varphi$ denotes the angle
between the transmission axis of the PBS and the $x$-axis. Then the field
amplitudes $\hat{a}_{1}$ and $\hat{a}_{2}$ of the transmitted and
reflected waves after the PBS can be written as
\begin{eqnarray}
     \hat{a}_{1}&=&e^{i\delta_x}(\hat{a}_x\cos{\varphi}+\hat{a}_y
     e^{-i\theta}\sin{\varphi}),
           \nonumber \\
     \hat{a}_{2}&=&e^{i\delta_x}(-\hat{a}_x\sin{\varphi}+\hat{a}_y
     e^{-i\theta}\cos{\varphi}),
           \label{defaij}
\end{eqnarray}
where $\theta=\delta_x-\delta_y$ is the phase difference between the $x$-
and $y$-components introduced by the compensator.

The secondary waves after PBS are photodetected and one observes the mean
values of the photocurrents $\langle i_1\rangle=\eta c \langle
\hat{a}^{\dag}_{1}\hat{a}_{1}\rangle$, and $\langle i_2\rangle=\eta c
\langle \hat{a}^{\dag}_{2}\hat{a}_{2}\rangle$, where $\eta$ is the quantum
efficiency of photodetection, and $c$ is the velocity of light (we have
put the charge of electron equal to unity so that the photocurrents are
measured in number of electrons per second). For simplicity in what
follows we shall consider the situation of $\eta=1$. Using
Eq.~(\ref{defaij}) we can write the mean photocurrent $\langle i_1\rangle$
measured in the transmission branch of the PBS as
\begin{equation}
     \langle i_1\rangle \equiv \langle i_1(\varphi,\theta)\rangle=
     \frac{1}{2}\eta c\Bigl[S_0+S_1\cos{2\varphi}+
     (S_2\cos{\theta}+S_3\sin{\theta})\sin{2\varphi}\Bigr],
            \label{photocurrent1}
\end{equation}
where $S_{\mu}$ are the classical Stokes parameters.

Equation (\ref{photocurrent1}) is the well-known formula for measuring the
four classical Stokes parameters. The first three of them are obtained by
removing the compensator $(\theta=0)$ and rotating the transmission axis
of the PBS to the angles $\varphi= 0^{\circ}, 45^{\circ}$, and
$90^{\circ}$, respectively. The fourth parameter, $S_3$, is measured by
using a compensator with $\theta=90^{\circ}$ or so-called quarter-wave
plate, and setting the transmission axis of the PBS to $\varphi=
45^{\circ}$. The four photocurrents are found to be, respectively,
\begin{eqnarray}
     \langle i_1(0^{\circ},0^{\circ})\rangle &=&
     \frac{1}{2}\eta c\left(S_0+S_1\right),
         \nonumber \\
     \langle i_1(45^{\circ},0^{\circ})\rangle &=&
     \frac{1}{2}\eta c\left(S_0+S_2\right),
         \nonumber \\
     \langle i_1(90^{\circ},0^{\circ})\rangle &=&
     \frac{1}{2}\eta c\left(S_0-S_1\right),
         \nonumber \\
     \langle i_1(45^{\circ},90^{\circ})\rangle &=&
     \frac{1}{2}\eta c\left(S_0+S_3\right).
         \label{class_Stokes}
\end{eqnarray}
Solving Eq.~(\ref{class_Stokes}) for $S_{\mu}$ we can obtain all classical
Stokes parameters from these four measurements.

\subsection{Observation of the fluctuation spectra of the quantum Stokes
parameters}

In quantum optics in addition to the mean values of the quantum Stokes
parameters $\langle \hat{S}_{\mu}\rangle$ their quantum fluctuations are
also taken into account. In this paper to describe the quantum fluctuation
we shall introduce the fluctuation spectra of the quantum Stokes
parameters.

Let us split the quantum Stokes operators $\hat{S}_{\mu}(t)$ given by
Eq.~(\ref{StokesS}) into the stationary mean value
$S_{\mu}=\langle\hat{S}_{\mu}\rangle$ and small fluctuation $\delta
\hat{S}_{\mu}(t)$,
\begin{equation}
     \hat{S}_{\mu}(t)=S_{\mu}+\delta \hat{S}_{\mu}(t).
\end{equation}
Taking the Fourier transform of $\delta \hat{S}_{\mu}(t)$,
\begin{equation}
     \delta \hat{S}_{\mu}(\Omega)=\frac{1}{\sqrt{2\pi}}\int\limits_{-\infty }
     ^{+\infty }\delta \hat{S}_{\mu}(t)e^{i\Omega t}dt,
             \label{fourier_Stokes}
\end{equation}
we can introduce the normally-ordered spectral correlation functions of
the fluctuations $\delta \hat{S}_{\mu}(\Omega)$ similar to the spectral
correlation functions of the quadrature components in
Eq.~(\ref{spectrquadii}), namely,
\begin{eqnarray}
      \langle :\delta \hat{S}_{\mu}(\Omega)\delta \hat{S}_{\mu}(\Omega'):
      \rangle &=&(\delta S_{\mu}^2)_{\Omega}\delta(\Omega+\Omega'),
               \nonumber\\
      \langle :\delta \hat{S}_{\mu}(\Omega)\delta \hat{S}_{\nu}(\Omega'):
      \rangle &=&(\delta S_{\mu}\delta S_{\nu})_{\Omega}\delta(\Omega+\Omega'),
      \quad(\mu\neq\nu).
               \label{spectrS}
\end{eqnarray}
Here $(\delta S_{\mu}^2)_{\Omega}$ are the spectral densities of the
corresponding fluctuations and $(\delta S_{\mu}\delta S_{\nu})_{\Omega}$
their cross-spectral densities. The symbol $:\dots:$ means normal ordering
of operators.

To measure the spectral densities $(\delta S_{\mu}^2)_{\Omega}$ and the
cross-spectral densities $(\delta S_{\mu}\delta S_{\nu})_{\Omega}$ of the
quantum Stokes parameters given by Eq.~(\ref{spectrS}) we can use an
experimental setup similar to one that we have used for the measurement of
the classical Stokes parameters (see Fig.~3). The difference is that
instead of detecting the mean photocurrents $\langle i_1\rangle$ and
$\langle i_2\rangle$ after the PBS, one observes now the photocurrent
fluctuation spectra $(\delta i^2_p)_{\Omega}, p=1,2$ defined as
\begin{equation}
     (\delta i^2_p)_{\Omega}=\int\limits_{-\infty}
     ^{+\infty}dt\; e^{i\Omega t}\langle \delta i_p(0)\delta i_p(t)\rangle,
                \label{spectrum_p}
\end{equation}
where $\langle \delta i_p(0)\delta i_p(t)\rangle$ is the correlation
function of the photocurrent fluctuations $\delta i_p(t)=i_p-\langle
i_p\rangle$, and $\langle i_p\rangle$ is the mean value of the
photocurrent. Alternatively, one can add and subtract the individual
photocurrents in the secondary channels and to investigate the sum
$i_{+}(t)=i_1(t)+i_2(t)$ and the difference $i_{-}(t)=i_1(t)-i_2(t)$ of
two photocurrents. In this case the information about the fluctuation
spectra of the quantum Stokes parameters is contained in the fluctuation
spectra
\begin{equation}
     (\delta i^2_{\pm})_{\Omega}=\int\limits_{-\infty}
     ^{+\infty}dt\; e^{i\Omega t}\langle \delta i_{\pm}(0)\delta
     i_{\pm}(t)\rangle.
                \label{spectrum_pm}
\end{equation}
The photocurrent fluctuation spectra $(\delta i^2_p)_{\Omega}$ and
$(\delta i^2_{\pm})_{\Omega}$ can be easily expressed through the spectral
densities $(\delta S^2_{\mu})_{\Omega}$ and the cross-spectral densities
$(\delta S_{\mu}\delta S_{\nu})_{\Omega}$ of the four quantum Stokes
parameters. The results are conveniently presented in terms of the
following linear combination of the three Stokes operators, $\hat{S}_1$,
$\hat{S}_2$, and $\hat{S}_3$,
\begin{equation}
      \hat{S}=\hat{S}_1\cos{2\varphi}+(\hat{S}_2\cos{\theta}+
      \hat{S}_3\sin{\theta})\sin{2\varphi},
               \label{pol_obs}
\end{equation}
which is sometimes called a polarization
observable~\cite{Karasev93,Buchev01}. We obtain the following expressions
for the fluctuation spectra $(\delta i^2_p)_{\Omega}$ and $(\delta
i^2_{\pm})_{\Omega}$, normalized to the shot-noise levels,
\begin{eqnarray}
     (\delta i^2_1)_{\Omega}/\langle i_1\rangle
     &=& 1+\frac{\kappa}{2\langle n_1\rangle}\Bigl[(\delta S^2_0)_{\Omega}+
     2(\delta S_0\delta S)_{\Omega}+(\delta S^2)_{\Omega}\Bigr],
            \label{spectr1} \\
     (\delta i^2_2)_{\Omega}/\langle i_2\rangle
     &=& 1+\frac{\kappa}{2\langle n_2\rangle}\Bigl[(\delta S^2_0)_{\Omega}-
     2(\delta S_0\delta S)_{\Omega}+(\delta S^2)_{\Omega}\Bigr],
            \label{spectr2} \\
     (\delta i^2_{-})_{\Omega}/\langle i_{+}\rangle
     &=& 1+\frac{2\kappa}{\langle n\rangle}(\delta S^2)_{\Omega},
            \label{spectr3} \\
     (\delta i^2_{+})_{\Omega}/\langle i_{+}\rangle
     &=& 1+\frac{2\kappa}{\langle n\rangle}(\delta S_0^2)_{\Omega},
            \label{spectr4}
\end{eqnarray}
where the corresponding spectral densities and cross-spectral densities of
are defined according to Eq.~(\ref{spectrS}). Here $\langle
i_+\rangle=\langle i_1\rangle+\langle i_2\rangle$ is the shot-noise level
of the photocurrent sum and difference, $\langle n_1\rangle=\langle
\hat{a}^{\dag}_{1}\hat{a}_{1}\rangle$, and $\langle n_2\rangle=\langle
\hat{a}^{\dag}_{2}\hat{a}_{2}\rangle$ are the mean photon numbers in the
corresponding secondary channels after the PBS, and $\langle
n\rangle=\langle n_1\rangle+\langle n_2\rangle$.

Equations (\ref{spectr1})-(\ref{spectr4}) are analogous of
Eq.~(\ref{photocurrent1}) for measuring the spectral densities of the
quantum Stokes parameters. It is clear from these equations that with
proper choice of angles $\theta$ and $\varphi$ all nonzero spectral
densities and cross-spectral densities of the Stokes operators can be
measured.

\subsection{Relations between the spectral densities of the quantum
Stokes parameters and of the quadrature components}

In Sec.~II D we have provided analytical results for the fluctuations of
the quadrature components $\delta X_x(\Omega)$, $\delta X_y(\Omega)$,
$\delta Y_y(\Omega)$, and for their spectral densities and cross-spectral
densities [see Esq.~(\ref{X_x})-(\ref{XY_y})]. Now we shall express the
spectral densities of the quantum Stokes operators through the spectral
densities of these quadrature components. As before, we shall restrict
ourselves to the case of the $x$-polarized stationary solution when
$\langle n_x\rangle = 2Q^2$ and $\langle n_y\rangle = 0.$

Using the same normal rule of correspondence between the operators and
their $c$-number representations as in Ref.~\cite{Hermier02} we shall
introduce the $c$-number variables $S_{\mu}(t)$ corresponding to the
quantum Stokes operators $\hat{S}_{\mu}(t)$. Since in Eq.~(\ref{StokesS})
the Stokes operators are normally ordered, the same relation holds true
for $S_{\mu}(t)$ and the $c$-number variables $a_i(t)$ and
$a^{\ast}_i(t)$, $i=x,y$.

Linearizing the $c$-number variables $S_{\mu}(t)$ around their stationary
values $S_{\mu}$ as
\begin{equation}
     S_{\mu}(t)=S_{\mu}+\delta S_{\mu}(t),
\end{equation}
we can express the fluctuations $\delta S_{\mu}(t)$ through the
fluctuations of the field components $\delta a_x(t)$ and $\delta a_y(t)$,
\begin{eqnarray}
     \delta S_0(t)&=&\delta S_1(t)=\sqrt{2} Q\Bigl(\delta a_x(t)+
     \delta a_x^{\ast}(t)\Bigr),
            \nonumber \\
     \delta S_2(t)&=&\sqrt{2} Q\Bigl(\delta a_y(t)+\delta a_y^{\ast}(t)\Bigr),
             \nonumber \\
     \delta S_3(t)&=&-\sqrt{2} i Q\Bigl(\delta a_y(t)-\delta
     a_y^{\ast}(t)\Bigr).
             \label{delta_S}
\end{eqnarray}
Taking into account Eq.~(\ref{quadratures}) we obtain the following
results relating the spectral densities of the Stokes operators with those
of the quadrature components,
\begin{eqnarray}
      (\delta S_0^2)_{\Omega}=(\delta S_1^2)_{\Omega} &=& 8Q^2 (\delta X_x^2)_{\Omega},
           \nonumber \\
      (\delta S_2^2)_{\Omega} &=& 8Q^2 (\delta X_y^2)_{\Omega},
           \nonumber \\
      (\delta S_3^2)_{\Omega} &=& 8Q^2 (\delta Y_y^2)_{\Omega},
           \nonumber \\
      (\delta S_2\delta S_3)_{\Omega} &=& 8Q^2 (\delta X_y\delta Y_y)_{\Omega}.
           \label{S_2S_3}
\end{eqnarray}
With help of these relations we arrive at,
\begin{eqnarray}
     (\delta i^2_1)_{\Omega}/\langle i_1\rangle
     &=& 1+8\kappa\Bigl[\cos^2{\varphi}(\delta X^2_x)_{\Omega}+
     \sin^2{\varphi}(\delta X^2_{\theta})_{\Omega}\Bigr],
            \label{spectrum1} \\
     (\delta i^2_2)_{\Omega}/\langle i_2\rangle
     &=& 1+8\kappa\Bigl[\sin^2{\varphi}(\delta X^2_x)_{\Omega}+
     \cos^2{\varphi}(\delta X^2_{\theta})_{\Omega}\Bigr],
            \label{spectrum2} \\
     (\delta i^2_{-})_{\Omega}/\langle i_{+}\rangle
     &=& 1+8\kappa\Bigl[\cos^2{2\varphi}(\delta X^2_x)_{\Omega}+
     \sin^2{2\varphi}(\delta X^2_{\theta})_{\Omega}\Bigr],
            \label{spectrum3} \\
     (\delta i^2_{+})_{\Omega}/\langle i_{+}\rangle
     &=& 1+8\kappa(\delta X^2_x)_{\Omega}.
            \label{spectrum4}
\end{eqnarray}
To simplify Eqs.~(\ref{spectrum1})-(\ref{spectrum3}) we have introduced
the following shorthand notation,
\begin{equation}
     \delta X_{\theta}(\Omega)=\cos{\theta}\;\delta X_y(\Omega)-
     \sin{\theta}\;\delta Y_y(\Omega),
            \label{X_theta}
\end{equation}
with its spectral density $(\delta X^2_{\theta})_{\Omega}$ given by,
\begin{equation}
     (\delta X^2_{\theta})_{\Omega}=\cos^2{\theta}(\delta X^2_y)_{\Omega}
     -2\sin{\theta}\cos{\theta}(\delta X_y\delta Y_y)_{\Omega}
     +\sin^2{\theta}(\delta Y^2_y)_{\Omega}.
            \label{spectrumX_theta}
\end{equation}
The mean values of the individual photocurrents $\langle i_1\rangle$ and
$\langle i_2\rangle$, and of the photocurrent sum $\langle
i_+\rangle=\langle i_1\rangle +\langle i_2\rangle$ are equal to
\begin{equation}
     \langle i_1\rangle=2Q^2\kappa\cos^2{\varphi}, \quad
     \langle i_2\rangle=2Q^2\kappa\sin^2{\varphi}, \quad
     \langle i_+\rangle=2Q^2\kappa.
            \label{mean_currents}
\end{equation}
In the next section we shall investigate in detail the spectral densities
of the quantum Stokes parameters and their cross-spectral densities.

\section{Polarization states of light in VCSELs}
\label{pol_states}
\subsection{Polarization squeezing}

The spectral densities $(\delta S_{\mu}^2)_{\Omega}$ of the quantum Stokes
parameters can be measured using any of three
Eqs.~(\ref{spectr1})-(\ref{spectr3}). Here we shall use
Eq.~(\ref{spectr3}) corresponding to observation of the noise spectrum
$(\delta i^2_{-})_{\Omega}(\varphi,\theta)$ of the photocurrent
difference. With help of Eq.~(\ref{pol_obs}) we can bring the photocurrent
noise spectrum $(\delta i^2_{-})_{\Omega}(\varphi,\theta)$ to the form
\begin{eqnarray}
     (\delta i^2_{-})_{\Omega}(\varphi,\theta)/\langle i_{+}\rangle
     &=& 1+\frac{2\kappa}{Q^2}\Bigl\{(\delta S_1^2)_{\Omega}\cos^2{2\varphi}+
     \sin^2{2\varphi}\Bigl[(\delta S_2^2)_{\Omega}\cos^2{\theta}
                \nonumber \\
     &-& (\delta S_2\delta S_3)_{\Omega}\;2\sin{\theta}\cos{\theta}
     +(\delta S_3^2)_{\Omega}\sin^2{\theta}\Bigr]\Bigr\}.
            \label{spectrum_P}
\end{eqnarray}
In this equation we have explicitly indicated the dependence of the
observed noise spectrum on the angle $\theta$ introduced by the
compensator and the angle $\varphi$ of the polarization beam splitter.

The spectral densities $(\delta S_0^2)_{\Omega}=(\delta S_1^2)_{\Omega}$
and $(\delta S_2^2)_{\Omega}$ of the  Stokes parameters $S_0, S_1$ and
$S_2$ are measured by removing the compensator $(\theta=0)$ and setting
the transmission axis of the PBS to the angles $\varphi=0^{\circ}$ and
$\varphi=45^{\circ}$. The spectral density of the parameter $S_3$ is
obtained by using a compensator with $\theta=90^{\circ}$ (quarter-wave
plate), and setting $\varphi=45^{\circ}$. The corresponding photocurrent
fluctuation spectra are given by,
\begin{eqnarray}
     (\delta i^2_{-})_{\Omega}(0^{\circ},0^{\circ})/\langle i_{+}\rangle
     &=& 1+\frac{2\kappa}{Q^2}(\delta S_1^2)_{\Omega},
                \label{Stokes1} \\
     (\delta i^2_{-})_{\Omega}(45^{\circ},0^{\circ})/\langle i_{+}\rangle
     &=& 1+\frac{2\kappa}{Q^2}(\delta S_2^2)_{\Omega},
                \label{Stokes2} \\
     (\delta i^2_{-})_{\Omega}(45^{\circ},90^{\circ})/\langle i_{+}\rangle
     &=& 1+\frac{2\kappa}{Q^2}(\delta S_3^2)_{\Omega},
                \label{Stokes3}
\end{eqnarray}
In Fig.~4 we have shown the photocurrent fluctuation spectra given by
Eqs.~(\ref{Stokes1})-(\ref{Stokes3}) for physical parameters close to that
used in experiment~\cite{Hermier02}, namely, $\kappa=100\;GHz$,
$\gamma=1\;GHz$, $\gamma_{\perp}=1000\;GHz$,$\gamma_{s}=50\;GHz$,
$\omega_{p}=40\;GHz$, $\alpha=-3$, $r=6$, and $p=1$. The parameter
$\kappa_a$, describing the dichroism of the laser crystal, was set equal
to zero in Fig.~4a, to $\kappa_a=10\;GHz$ in Fig.~4b and to
$\kappa_a=50\;GHz$ in Fig.~4c.

Let us first discuss the case without dichroism (Fig.~4a). As seen from
Fig.~4a, the spectral density $(\delta S_1^2)_{\Omega}$ of the Stokes
parameter $S_1$ has a peak at a characteristic frequency $\Omega_1$, while
two other spectra $(\delta S_2^2)_{\Omega}$ and $(\delta S_3^2)_{\Omega}$
for the Stokes parameters $S_2$ and $S_3$ exhibit peaks at another
(higher) characteristic frequency $\Omega_2$. These peaks are well-known
from the theory of solid-state and semiconductor lasers and have their
physical origin in the relaxation oscillations due to a periodic energy
exchange between the active medium and the laser radiation. Since in our
case there are two upper levels $|a+\rangle$ and $|a-\rangle$ in the
active laser medium, we have two subsystems where the periodic energy
exchange takes place independently. First subsystem is described by the
total population $D$ of the upper levels and the Stokes parameter $S_1$
[see Eqs.~(\ref{polxtime})], and its frequency of the relaxation
oscillations is equal to $\Omega_1$. In the second subsystem the
relaxation oscillations take place between the population difference $d$
and the two Stokes parameters $S_2$ and $S_3$ at the frequency $\Omega_2$
[see Eqs.~(\ref{polytime})].

Second important feature that one can observe in Fig.~4a is reduction of
the quantum fluctuations of the Stokes parameter $S_1$ below the standard
quantum limit at low frequencies $\Omega$ in the case of regular pumping,
$p=1$. Thus, we can speak of phenomenon of {\it polarization squeezing}
with respect to $S_1$ in VCSELs with regular pumping. This result is to be
expected. In fact, as follows from Eqs.~(\ref{StokesS}), for the
$x$-polarized stationary solution the Stokes parameter $S_1$ coincides
with the total number of photons in the laser field. It is well known from
the literature~\cite{Golubev} that a regularly pumped two-level laser can
exhibit the sub-Poissonian photon statistics, i.~e.~the fluctuations of
its photon number could be reduced below the standard quantum limit. One
could therefore say that the polarization squeezing with the respect to
$S_1$ in a regularly pumped VCSEL is the consequence of the sub-Poissonian
statistics of photons.

However, it is worth noting that the relation between the sub-Poissonian
statistics of photons and the regular pumping statistics in VCSELs is not
so direct as in the case of a two-level laser considered
in~\cite{Golubev}. Indeed, due to the degeneracy of the upper laser level
on two sublevels $|a+\rangle$ and $|a-\rangle$, the regular pumping of the
total population $D$ of the upper level remains random for each individual
sublevel due to the partition noise. It turns out that in the case of
$x$-polarized stationary solution this partition noise does not contribute
to the fluctuations of the total photon number and of the Stokes parameter
$S_1$. The reason for this is that, as follows from Eqs.~(\ref{polxtime}),
the fluctuations of the Stokes parameter $S_1$ are coupled only with the
fluctuations of the total population $D$ and not with fluctuations of the
populations of individual sublevels.

The role of dichroism is illustrated in Fig.~4b and 4c. As seen from these
figures, appearance of dichroism in the system has two major consequences.
Firstly, the quantum noise reduction below the standard quantum limit in
the spectral density $(\delta S_1^2)_{\Omega}$ of the first Stokes
parameter is deteriorated by the factor $\kappa/(\kappa+\kappa_a)$. This
deterioration has a clear physical explanation. Nonzero dichroism
introduces random losses of the laser radiation inside the resonator at
the rate $\kappa_a$. The total decay rate of the laser field inside the
resonator is now given by $\kappa+\kappa_a$, while the outcoupling rate
determined by the transmission of the cavity mirror is equal to $\kappa$.

The second consequence of dichroism in the system is suppression of the
relaxation oscillations at the frequency $\Omega_2$ related to the Stokes
parameters $S_2$ and $S_3$. We can see from Fig.~4b that for small values
of $\kappa_a$ ($\kappa_a=10\;GHz$ while $\kappa=100\;GHz$) the peak of
relaxation oscillations at $\Omega_2$ becomes more pronounced. This is
explained by the fact that for these values of $\kappa_a$ we approach
closer to the instability region. However, with increasing $\kappa_a$ as
in Fig.~4c the relaxation oscillations at $\Omega_2$ rapidly disappear.

The three spectral densities $(\delta S_1^2)_{\Omega}, (\delta
S_2^2)_{\Omega}$ and $(\delta S_1^2)_{\Omega}$ in Fig.~4 can be also
interpreted in terms of the uncertainty ellipsoid that we have mentioned
in Sec.~III A. Since the spectral densities depend on the frequency
$\Omega$, one has to speak about the frequency-dependent uncertainty
ellipsoid with tree major axis determined by the corresponding spectral
densities. These spectral densities are normalized to the shot-noise level
so that a sphere of unit radius in the Stokes-Poincar\'e space corresponds
to the standard quantum limit realized for a coherent polarization state.
As follows from Fig.~4a, for example, for a polarization-squeezed state in
the area of low frequencies, where $(\delta S_1^2)_{\Omega}$ is below the
standard quantum limit, the uncertainty ellipsoid has the shape of a
pancake. Instead, in the vicinity of the frequency of relaxation
oscillations $\Omega_1$ this uncertainty ellipsoid takes a cigar-like
shape with $(\delta S_1^2)_{\Omega}$ larger than two other components.

\subsection{Cross-correlation spectrum of photocurrents}

Using the experimental setup shown in Fig.~3 one can also measure the
cross-correlation function of fluctuations between the photocurrents
$i_1(t)$ and $i_2(t)$, i.~e.~$\langle \delta i_1(0)\delta i_2(t)\rangle$,
or the corresponding cross-correlation spectrum of fluctuations,
\begin{equation}
     (\delta i_1\delta i_2)_{\Omega}=\int\limits_{-\infty}
     ^{+\infty}dt\; e^{i\Omega t}\langle \delta i_1(0)\delta
     i_2(t)\rangle.
                \label{cross-spectrum}
\end{equation}
Usually it is more customary to work with the normalized cross-correlation
spectrum of the photocurrent fluctuations,
\begin{equation}
     C_{12}(\Omega)=\frac{(\delta i_1\delta i_2)_{\Omega}}
     {\sqrt{(\delta i_1^2)_{\Omega}}\sqrt{(\delta i_2^2)_{\Omega}}}.
            \label{cross-corr}
\end{equation}
Using the Cauchy-Schwartz inequality one can demonstrate that this
spectrum is normalized as $|C_{12}(\Omega)|\leq 1$. Hence,
$C_{12}(\Omega)=-1$ corresponds to the maximum anticorrelations between
the two photocurrents, while $C_{12}(\Omega)=1$ to the maximum
correlations. Experimentally this spectrum can be measured as,
\begin{equation}
     C_{12}(\Omega)=\frac{(\delta i_+^2)_{\Omega}-(\delta i_1^2)_{\Omega}-
     (\delta i_2^2)_{\Omega}}
     {2\sqrt{(\delta i_1^2)_{\Omega}(\delta i_2^2)_{\Omega}}}.
\end{equation}
The normalized cross-correlation spectrum $C_{12}(\Omega)$ can be
expressed through the spectral densities and cross-spectral densities of
the amplitude quadrature components $\delta X_1$ and $\delta X_2$ as,
\begin{equation}
     C_{12}(\Omega)=\frac{8\kappa(\delta X_1\delta X_2)_{\Omega}}
     {\sqrt{1+8\kappa(\delta X^2_1)_{\Omega}}
     \sqrt{1+8\kappa(\delta X^2_2)_{\Omega}}}.
              \label{cross}
\end{equation}
Using the relations between the field amplitudes $\hat{a}_1$ and
$\hat{a}_2$ of the transmitted and reflected waves after the PBS and the
incoming amplitudes $\hat{a}_x$ and $\hat{a}_y$, given by
Eq.~(\ref{defaij}), we obtain
\begin{eqnarray}
     (\delta X_1\delta X_2)_{\Omega} &=& \cos{\varphi} \sin{\varphi}
     \Bigl[(\delta X_x^2)_{\Omega}-(\delta X_{\theta}^2)_{\Omega}\Bigr],
            \nonumber \\
     (\delta X^2_1)_{\Omega}&=& \cos^2{\varphi}(\delta X^2_x)_{\Omega}+
     \sin^2{\varphi}(\delta X^2_{\theta})_{\Omega},
            \nonumber \\
     (\delta X^2_2)_{\Omega}&=& \sin^2{\varphi}(\delta X^2_x)_{\Omega}+
     \cos^2{\varphi}(\delta X^2_{\theta})_{\Omega}.
            \label{22}
\end{eqnarray}
These relations allow us to express the cross-correlation spectrum
$C_{12}(\Omega)$ in terms of the spectral densities $(\delta
X^2_x)_{\Omega}$ and $(\delta X^2_{\theta})_{\Omega}$ calculated earlier.

In Fig.~5 we have plotted the cross-correlation spectrum $C_{12}(\Omega)$
for $\varphi=\pi/4$ and $\theta=0$. In this case the general result for
$C_{12}(\Omega)$ given by Eqs.~(\ref{cross})-(\ref{22}) is simplified to,
\begin{equation}
     C_{12}(\Omega)=\frac{4\kappa\Bigl[(\delta X_x^2)_{\Omega}-
     (\delta X_y^2)_{\Omega}\Bigr]}{1+4\kappa\Bigl[(\delta
     X_x^2)_{\Omega}+(\delta X_y^2)_{\Omega}\Bigr]}.
              \label{cross_simpl}
\end{equation}
Fig.~5a shows this cross-correlation spectrum for the case without
dichroism and the same values of physical parameters as in Fig.~4. As
follows from Fig.~5a, the cross-correlations are absent at high
frequencies $\Omega$ larger than $30\;GHz$. At lower frequencies of the
order of $15\;GHz$ the curve of $C_{12}(\Omega)$ shows anticorrelations
which turn to correlations at still lower frequencies of the order of
$5\;GHz$. In the area of low frequencies $\Omega$ smaller then $1\;GHz$
one has again anticorrelations.

This oscillating behavior of the cross-correlation spectrum
$C_{12}(\Omega)$ is in full agreement with behavior of the fluctuation
spectra of the Stokes parameters $S_1$ and $S_2$ in Fig.~4a. Indeed, the
cross-correlation function $C_{12}(\Omega)$ is proportional to the
difference of the spectral densities of the quadrature components $(\delta
X_x^2)_{\Omega}-(\delta X_y^2)_{\Omega}$ [or the corresponding Stokes
parameters, $(\delta S_1^2)_{\Omega}-(\delta S_2^2)_{\Omega}$]. Therefore,
for $(\delta X_x^2)_{\Omega}>(\delta X_y^2)_{\Omega}$ we have correlations
between the two photocurrents, while in the opposite case -
anticorrelations.

Fig.~5b illustrates the same cross-correlation spectrum in presence of
dichroism for different values of parameter $\kappa_a$. As mentioned
above, the essential role of dichroism is in the suppression of the
relaxation oscillations. When $\kappa_a$ approaches the critical value
$\kappa_a= 10\;GHz$ of the instability border, the relaxation oscillations
grow up and reinforce anticorrelations. Further increase of $\kappa_a$
results in suppression of the relaxation oscillations and respectively in
transformation of anticorrelations into correlations for $\kappa_a$ larger
than $50\;GHz$.

\subsection{Cross-correlations between the Stokes parameters $S_2$ and $S_3$}

For the $x$-polarized stationary solution that we consider in this paper,
the linearized field operator $\vec{\hat{E}}(t)$ from
Eq.~(\ref{vector_field}) can be approximately written as,
\begin{equation}
     \vec{\hat{E}}(t)=e^{i\Delta t}\Bigl[\sqrt{2}Q+\delta\hat{X}_x(t)+
     i\delta\hat{Y}_x(t)\Bigr]\Bigl[\vec{e}_x+\frac{1}{\sqrt{2}Q}
     (\delta\hat{X}_y(t)+i\delta\hat{Y}_y(t))\vec{e}_{y}\Bigr].
           \label{vector_field_2}
\end{equation}
This representation of the linearized field operator is very useful as it
clarifies the physical meaning of the quantum fluctuations of the four
quadrature components that appear in Eq.~(\ref{vector_field_2}). The
fluctuations $\delta\hat{X}_x(t)$ and $\delta\hat{Y}_x(t)$ describe
respectively the quantum fluctuations of the amplitude and the phase of
the electromagnetic field $\vec{\hat{E}}(t)$. The quantum fluctuations of
two other quadrature components $\delta\hat{X}_y(t)$ and
$\delta\hat{Y}_y(t)$ characterize the quantum fluctuations of the {\it
polarization} of the field $\vec{\hat{E}}(t)$. To see this more clear let
us compare Eq.~(\ref{vector_field_2}) with the classical expression often
used in the literature on VCSELs (see for example Ref.~\cite{vanExter}),
\begin{equation}
     \vec{\hat{E}}(t)\approx e^{i\Delta t}|E|\Bigl[\vec{e}_x-(\delta\phi
     +i\delta\chi)\vec{e}_{y}\Bigr].
           \label{vector_field_class}
\end{equation}
In this expression we have neglected the amplitude and the phase
fluctuations of the field and have introduced the fluctuations
$\delta\phi$ and $\delta\chi$, $\delta\phi\ll 1, \delta\chi\ll 1$ of two
angles $\phi$ and $\chi$, that characterize the optical polarization state
on the Poincar\'e sphere. The first angle $\phi\quad (0\leq\phi\leq \pi)$
is called the polarization angle and it determines the direction of the
polarization ellipse. The second angle $\chi\quad (-\pi/4\leq\chi\leq
\pi/4)$ is the ellipticity angle. For $x$-polarized field both of these
angles are zero. Comparing Eq.~(\ref{vector_field_2}) and
Eq.~(\ref{vector_field_class}) we conclude that these two classical
fluctuations can be associated with their quantum counterparts as
$\displaystyle{\delta\phi\rightarrow -\frac{\delta\hat{X}_y}{\sqrt{2}Q}}$
and $\displaystyle{\delta\chi\rightarrow
-\frac{\delta\hat{Y}_y}{\sqrt{2}Q}}$. Taking into account
Eq.~(\ref{delta_S}) we can also write $\displaystyle{\delta\phi\rightarrow
-\frac{\delta\hat{S}_2}{4Q^2}}$ and $\displaystyle{\delta\chi\rightarrow
-\frac{\delta\hat{S}_3}{4Q^2}}$.

Thus, the quantum fluctuations of the Stokes parameter $S_2$ characterize
the fluctuations of the polarization angle, and those of the $S_3$ - the
fluctuations of the ellipticity angle. In the subsection A we have
evaluated the fluctuation spectra of the Stokes parameters $S_2$ and
$S_3$. However, as follows from Eq.~(\ref{S_2S_3}) these two parameters
are also cross-correlated. Hence, we shall introduce the cross-correlation
spectrum $C_{23}(\Omega)$ between these two parameters in the same way as
we did for characterization of the cross-correlations of two
photocurrents,
\begin{equation}
     C_{23}(\Omega)=\frac{(\delta S_2\delta S_3)_{\Omega}}
     {\sqrt{(\delta S_2^2)_{\Omega}}\sqrt{(\delta S_3^2)_{\Omega}}}.
            \label{cross-corr2}
\end{equation}
This cross-correlation spectrum is normalized as $|C_{23}(\Omega)|\leq 1$
and can be experimentally determined from the measurements of the
following three photocurrent fluctuation spectra,
\begin{eqnarray}
     (\delta i^2_{-})_{\Omega}(45^{\circ},0^{\circ})/\langle i_{+}\rangle
     &=& 1+\frac{2\kappa}{Q^2}(\delta S_2^2)_{\Omega},
                \label{Cross1} \\
     (\delta i^2_{-})_{\Omega}(45^{\circ},90^{\circ})/\langle i_{+}\rangle
     &=& 1+\frac{2\kappa}{Q^2}(\delta S_3^2)_{\Omega},
                \label{Cross2} \\
     (\delta i^2_{-})_{\Omega}(45^{\circ},45^{\circ})/\langle i_{+}\rangle
     &=& 1+\frac{\kappa}{Q^2}\Bigl[(\delta S_2^2)_{\Omega}+(\delta S_3^2)_{\Omega}+
     2(\delta S_2\delta S_3)_{\Omega}\Bigr].
                \label{Cross3}
\end{eqnarray}
We have numerically evaluated the cross-correlation spectrum
$C_{23}(\Omega)$ for the same values of physical parameters as in the
previous subsection. In Fig.~6 we illustrate these spectra in the absence
of dichroism $(\kappa_a=0)$ and for two different values of $\kappa_a$
equal to $10\;GHz$ and $50\;GHz$.

As follows from this figure, in the absence of dichroism the
cross-correlation spectrum shows negative correlations at low frequencies
$\Omega$ less than $10\;GHz$. These anticorrelations appear due to the
coupling between the Stokes parameters $S_2$ and $S_3$ via the population
difference $d$. For higher frequencies this coupling becomes less
efficient and for $\Omega$ higher than $30\;GHz$ the fluctuations of $S_2$
and $S_3$ become independent $(C_{23}\rightarrow 0)$.

For nonzero dichroism the anticorrelations between $S_2$ and $S_3$ at low
frequencies firstly disappear and then turn into positive correlations for
larger values of $\kappa_a$, for example at $\kappa_a=50\;GHz$. Thus,
dichroism changes the nature of correlations between $S_2$ and $S_3$.

\section{Conclusions}
\label{concl}

In conclusion we have presented a generalized and fully analytical theory
of quantum fluctuations in VCSELs, proposed for the first time in
Ref.~\cite{Hermier02}. The original results of our investigation are the
analytical expressions for the spectral densities of the quadrature field
components and of the corresponding quantum Stokes parameters. These
analytical results facilitate the comparison between the theory and the
experimental measurements. Moreover, we have included into the theory a
nonzero linear dichroism of the semiconductor medium that was neglected in
Ref.~\cite{Hermier02}.

Our theory is very closely related to possible experimental observation of
the quantum fluctuations in VCSEls that can be performed in a
correlation-type measurement shown in Fig.~3. We have calculated
analytically and illustrated graphically the typical fluctuation and
cross-correlation spectra that could be observed in this type of
measurements. Our theoretical results allow for direct comparison with
experiments.

We predict theoretically polarization squeezing in VCSELs when the quantum
fluctuations of the Stokes parameter $S_1$ are reduced below the standard
quantum limit. This phenomenon has its origin in regular pumping
statistics of the active laser medium. However, the regularity in the
pumping statistics alone is not sufficient for polarization squeezing in
this type of lasers due to the partition noise between two upper sublevels
in the laser medium. The second important feature of VCSELs that
guarantees polarization squeezing is their dynamical behavior that couples
the statistical properties of the Stokes parameter $S_1$ only with those
of the {\it total} population of two upper sublevels.

We have analyzed the role of linear dichroism and have concluded that it
mainly influences the relaxation oscillations in VCSEls. These
oscillations are typical for the solid-state and semiconductor lasers. The
particularity of VCSEls is that in this case there are two types of
relaxation oscillations with clearly distinct characteristic frequencies
$\Omega_1$ and $\Omega_2$. First oscillations (with frequency $\Omega_1$)
are related to the total population of two upper sub-levels and they
contribute to the fluctuation spectrum of the Stokes parameter $S_1$. The
second type of relaxation oscillations (with frequency $\Omega_2$) is
connected with the population difference and its peak appears in the
fluctuation spectra of the Stokes parameters $S_2$ and $S_3$. It turns out
the dichroism dumps the relaxation oscillations of the second type and
does not influence those of the first type. To understand this result let
us recall that the relaxation oscillations appear in the lasers of the
second type when the resonator losses are more rapid compared with those
of the laser medium. As follows from Eqs.~(\ref{polxtime}) and
(\ref{polytime}) dichroism increases the losses for the $y$-polarized
light component coupled with the population difference $d$ and does not
change those of the $x$-polarized component related to population sum $D$.

\begin{acknowledgments}
This work was performed within the Franco-Russian cooperation program
``Lasers and Advanced Optical Information Technologies'' with financial
support from the following organizations: INTAS (grant INTAS-01-2097),
RFBR (grant 03-02-16035), Minvuz of Russia (grant E 02-3.2-239), and by
the Russian program ``Universities of Russia'' (grant ur.01.01.041).
\end{acknowledgments}


 \newpage

 \begin{figure}[t]
 \includegraphics[width=120mm]{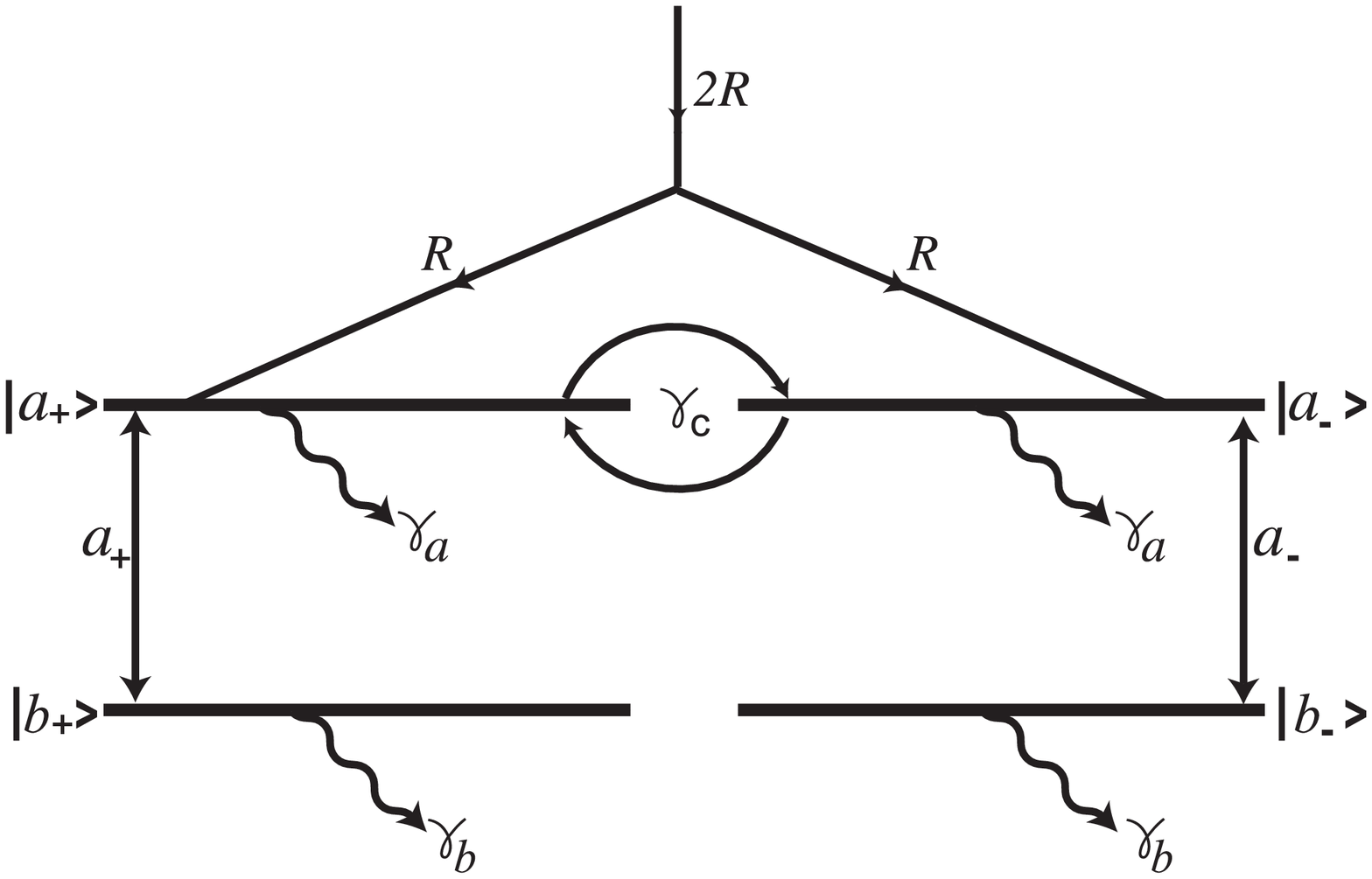}
 \caption{Four-level scheme of the active medium of VCSEL.}
 \label{fig1}
 \end{figure}

 \newpage

 \begin{figure}[t]
 \includegraphics[width=120mm]{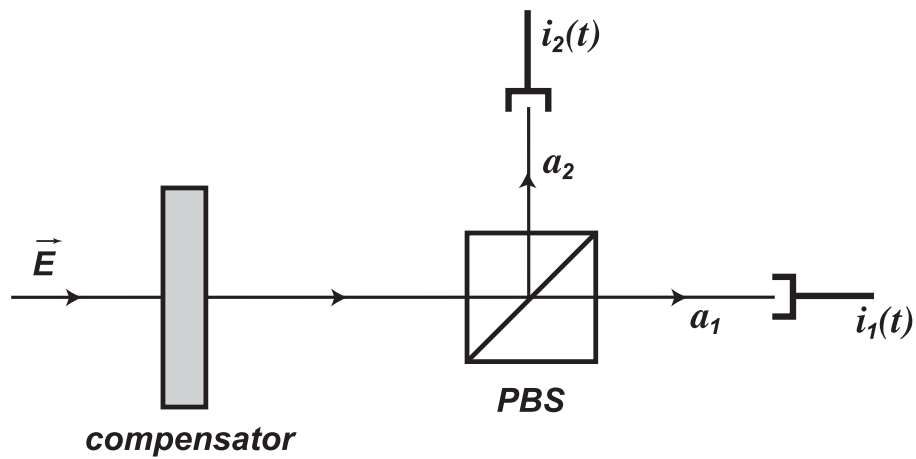}
 \caption{Experimental setup for measurement of the classical
 Stokes parameters.}
 \label{fig2}
 \end{figure}

 \begin{figure}[t]
 \includegraphics[width=120mm]{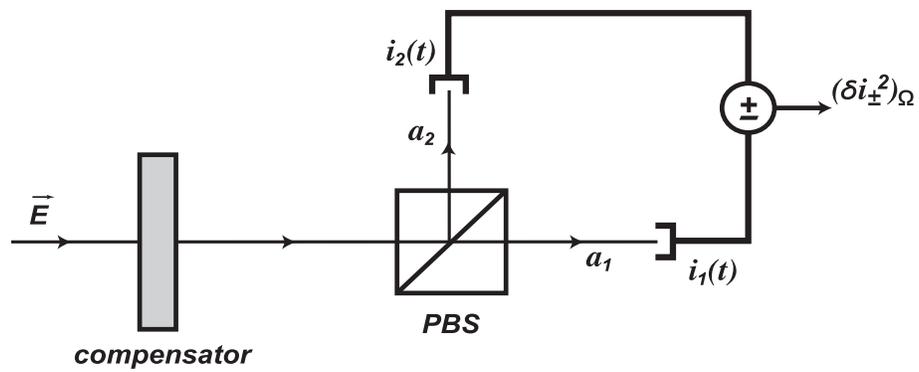}
 \caption{Experimental scheme for measurement of the spectral densities
 and cross-spectral densities of the quantum Stokes parameters.}
 \label{fig3}
 \end{figure}

 \begin{figure}[t]
 \includegraphics[width=120mm]{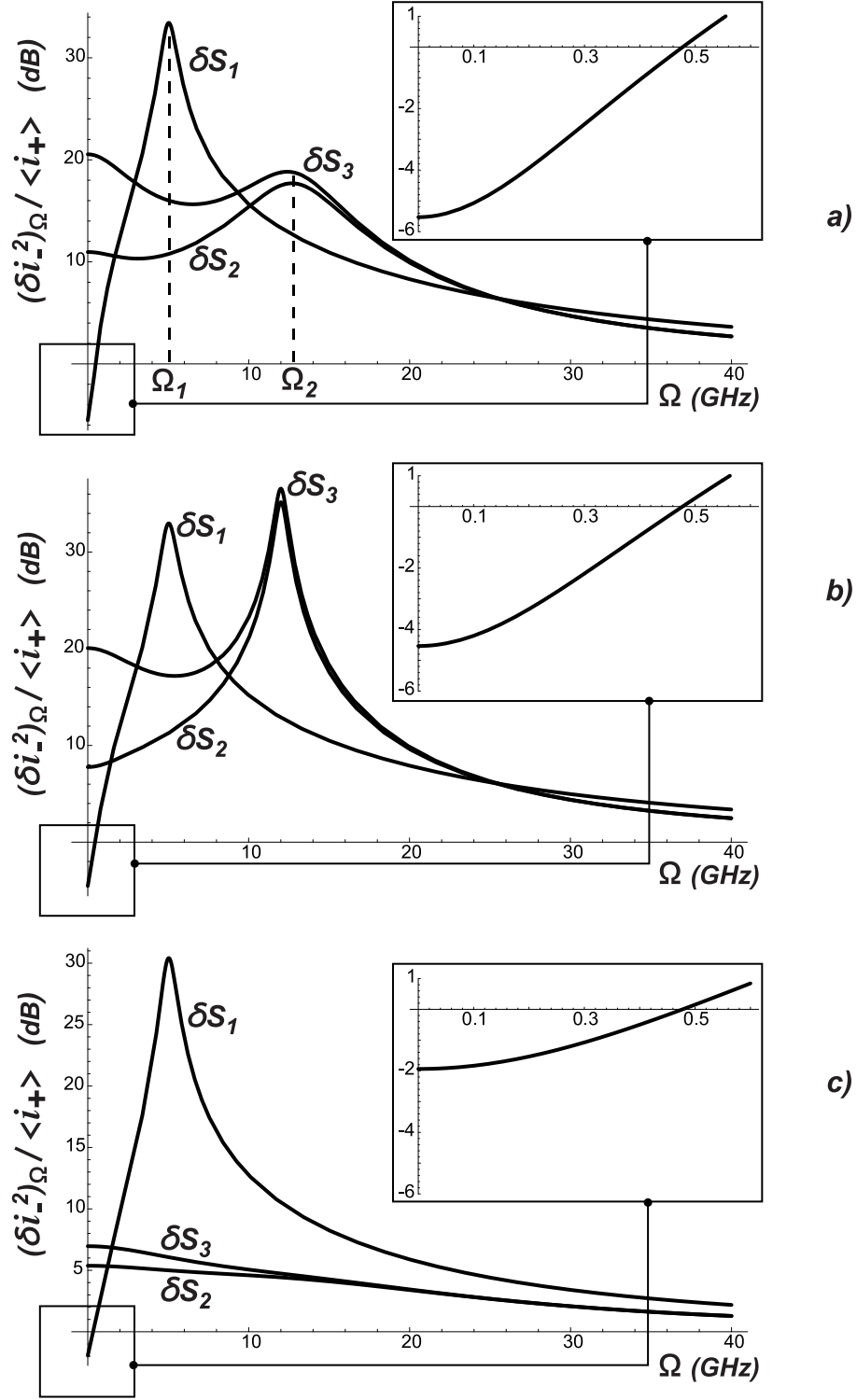}
 \caption{Photocurrent fluctuation spectra for the Stokes parameters
 $S_1, S_2$ and $S_3$; a) without dichroism, $\kappa_a=0$, b) with
 dichroism, $\kappa_a=10\;GHz$, and c) with $\kappa_a=50\;GHz$. The
 values of other parameters are:
 $\kappa=100\;GHz$, $\gamma=1\;GHz$, $\gamma_{\perp}=1000\;GHz$,
 $\gamma_{s}=50\;GHz$, $\omega_{p}=40\;GHz$, $\alpha=3$ and $p=1$.}
 \label{fig4}
 \end{figure}

 \begin{figure}[t]
 \includegraphics[width=120mm]{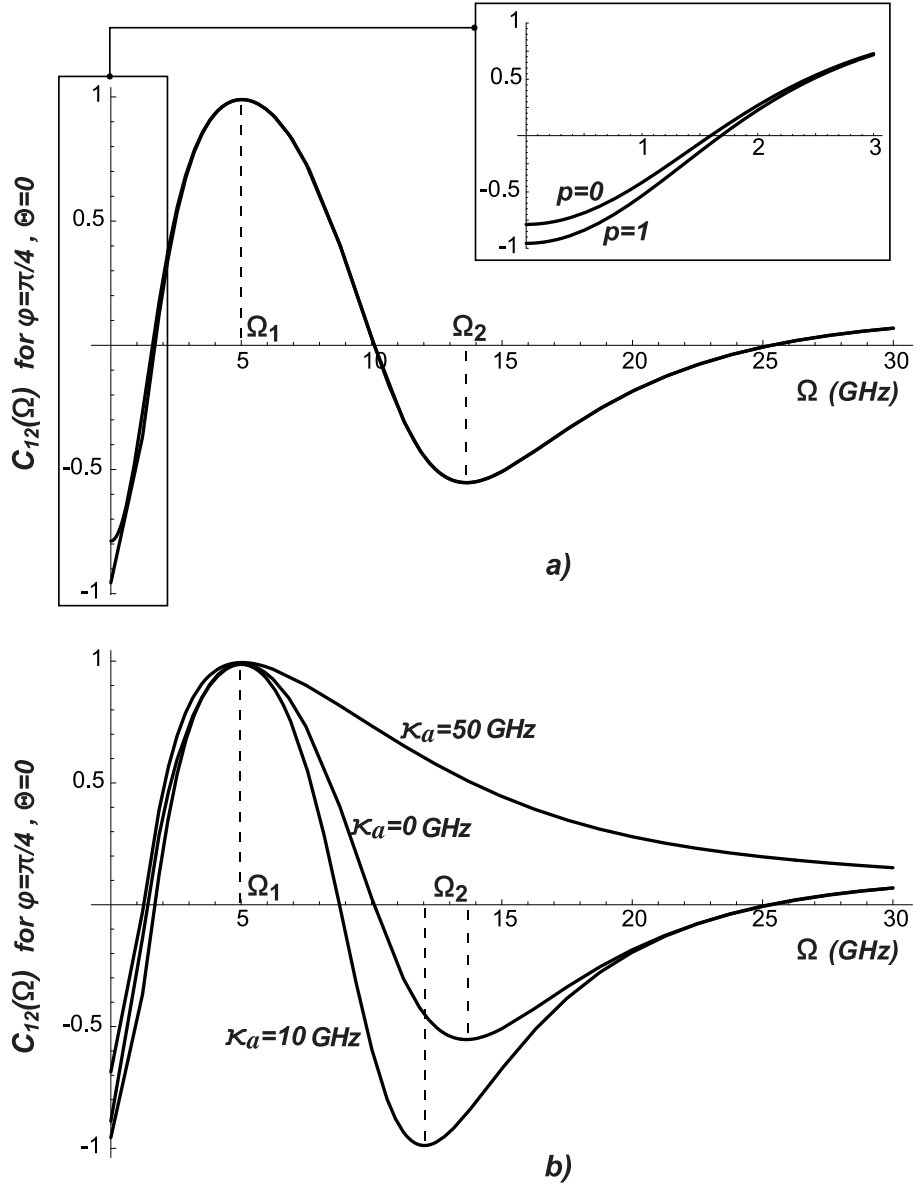}
 \caption{Cross-correlation spectrum $C_{12}(\Omega)$ for $\varphi=\pi/4$
 and $\theta=0$; a) without dichroism, $\kappa_a=0$ and b) with
 dichroism, $\kappa_a=10\;GHz$ and $\kappa_a=50\;GHz$. The inset in a)
 illustrates the role of the statistical parameter $p$ at low spectral
 frequencies. All other parameters are as in Fig.~4.}
 \label{fig5}
 \end{figure}

 \begin{figure}[t]
 \includegraphics[width=120mm]{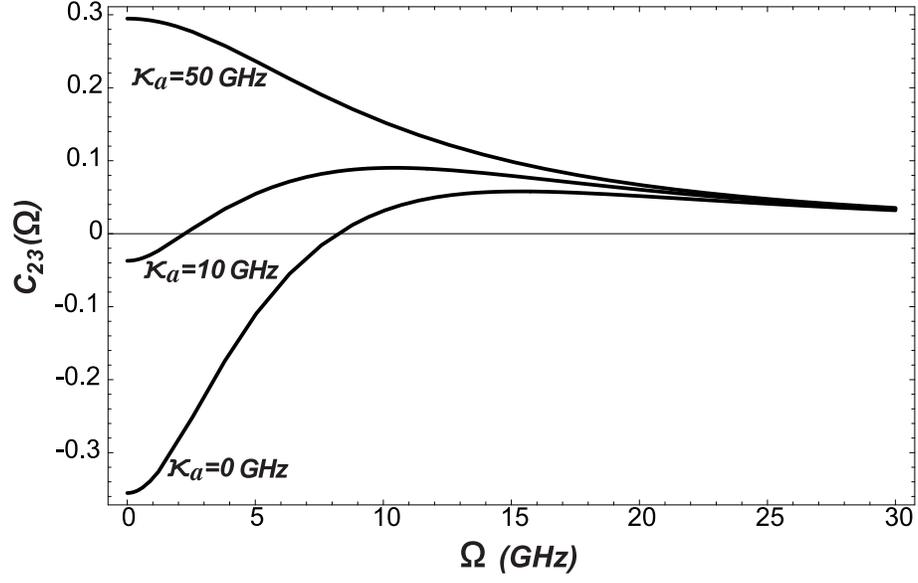}
 \caption{Cross-correlation spectrum $C_{23}(\Omega)$ without dichroism,
 $\kappa_a=0$ and with dichroism, $\kappa_a=10\;GHz$ and $\kappa_a=50\;GHz$
 for the same values of physical parameters as in Fig.~4.}
 \label{fig6}
 \end{figure}

\end{document}